\documentstyle[12pt]{article}
\def\be{\begin{equation}}
\def\ee{\end{equation}}

\begin{document}

\title{$p$--Adic representation of the Cuntz algebra \\ and the free coherent states}

\author{S.V.Kozyrev}

\maketitle

\begin{abstract}
Representation of the Cuntz algebra in the space of (complex
valued) functions on $p$--adic disk is introduced. The relation of
this representation and the free coherent states is investigated.
\end{abstract}

\section{Introduction}

The present paper combines the investigations on $p$--adic
mathematical physics and noncommutative probability. We introduce
the representation of the Cuntz algebra in the space of (complex
valued) functions on $p$--adic disk and investigate the relation
of this representation and the free coherent states. The
representation which we introduce will be unitarily equivalent to
one of the representation considered by Bratteli and Yorgensen in
\cite{BY} (without use of $p$--adic analysis).

Continuing the investigations of \cite{coherent1},
\cite{coherent2}, \cite{coherent3}, we investigate the free
coherent states (or shortly FCS), which are (unbounded)
eigenvectors of the linear combination of annihilators in the free
Fock space. In \cite{coherent1}, \cite{coherent2} it was shown
that the space of the free coherent states is highly degenerate
for the fixed eigenvalue $\lambda$ (and infinite dimensional), and
this degeneracy is naturally described by the space $D'(Z_p)$ of
generalized functions on $p$--adic disk ($p$ is a number of
independent creators in the free Fock space). In \cite{coherent3}
the results of \cite{coherent1}, \cite{coherent2} were
reformulated using the language of rigged Hilbert spaces and the
interpretation of the relation between the free coherent states
and $p$--adics using noncommutative geometry was proposed (see
also \cite{phase}).

$p$--Adic mathematical physics studies the problems of
mathematical physics with the help of $p$--adic analysis.
$p$--Adic mathematical physics was studied in
\cite{coherent1}--\cite{PaSu}. For instance in the book \cite{VVZ}
the analysis of $p$--adic pseudodifferential operators was
developed. In \cite{Vstring} a $p$--adic approach in the string
theory was proposed. In \cite{Khren} a theory of $p$--adic valued
distributions was investigated. In \cite{ABK}, \cite{PaSu} it was
shown that the Parisi matrix used in the replica method is
equivalent, in the simplest case, to a $p$--adic
pseudodifferential operator. In \cite{wavelets} it was shown that
the wavelet basis in $L^2(R)$ after the $p$--adic change of
variable (the continuous map of $p$--adic numbers onto real
numbers conserving the measure) maps onto the basis of
eigenvectors of the Vladimirov operator of $p$--adic fractional
derivation. In \cite{phase} a procedure to generate the
ultrametric space used in the replica approach was proposed.

The Free (or quantum Boltzmann) Fock space has been considered in
some works on quantum chromodynamics \cite{MasterFld} and
noncommutative probability \cite{Voi92}--\cite{qdeform}.

Discuss in short the results of \cite{coherent1},
\cite{coherent2}, \cite{coherent3}.

The free Fock space ${\cal F}$ over a Hilbert space  ${\cal H}$ is
the completion of the tensor algebra
$${\cal F}=\oplus_{n=0}^{\infty}{\cal H}^{\otimes n}.$$
Creation and annihilation operators act as follows:
$$
A^{\dag}(f) f_{1}\otimes \dots \otimes f_{n}=f\otimes f_{1}\otimes
\dots \otimes f_{n};\quad f,f_i \in {\cal H}
$$
$$
A(f) f_{1}\otimes \dots \otimes f_{n}=\langle f,f_{1} \rangle
f_{2}\otimes \dots  \otimes f_{n}; \quad f,f_i \in {\cal H}
$$
where  $\langle \cdot,\cdot \rangle $ is the scalar product in the
Hilbert space ${\cal H}$. Scalar product in the free Fock space
(which we also denote $\langle \cdot,\cdot \rangle $) is defined
in the standard way.

In the case when  ${\cal H}$ is the $p$--dimensional complex
Euclidean space we have $p$ creation operators $A^{\dag}_{i}$,
$i=0,\dots ,p-1$; $p$ annihilation operators $A_{i}$, $i=0,\dots
,p-1$ with the relations
\begin{equation}\label{aac}
A_{i}A_{j}^{\dag}=\delta_{ij}.
\end{equation}
and the vacuum vector $\Omega$  in the free Fock space satisfies
\begin{equation}\label{vacuum}
A_{i}\Omega=0.
\end{equation}

We will also use the following factor--algebra of the quantum
Boltzmann algebra. The Cuntz algebra (with $p$ degrees of freedom)
is the algebra with involution which is generated by $p$ creation
operators $A^{\dag}_{i}$, $i=0,\dots ,p-1$; $p$ annihilation
operators $A_{i}$, $i=0,\dots ,p-1$ with commutation relations
\begin{equation}\label{aac1}
A_{i}A_{j}^{\dag}=\delta_{ij};
\end{equation}
\begin{equation}\label{cuntz}
\sum_{i=0}^{p-1}A_{i}^{\dag}A_{i}=1.
\end{equation}

The free coherent states (or shortly FCS) were introduced in
\cite{coherent1}, \cite{coherent2} as the formal eigenvectors of
the annihilation operator $A=\sum_{i=0}^{p-1}A_{i}$ in the free
Fock space ${\cal F}$ for some eigenvalue $\lambda$,
\begin{equation}\label{freecoherent}
A \Psi= \lambda \Psi.
\end{equation}
The formal solution of (\ref{freecoherent}) is
\begin{equation}\label{psi}
\Psi=\sum_{I} \lambda^{|I|} \Psi_I A^{\dag}_{I}\Omega.
\end{equation}
Here  the multiindex $I=i_0 \dots i_{k-1}$, $i_j \in\{0, \dots
,p-1\}$ and \be\label{Adag} A^{\dag}_{I}=A^{\dag}_{i_{k-1}} \dots
A^{\dag}_{i_0} \ee $\Psi_I$ are complex numbers which satisfy
\begin{equation}\label{cascade}
\Psi_I=\sum_{i=0}^{p-1}\Psi_{Ii}.
\end{equation}
The summation in the formula  (\ref{psi}) runs on all sequences
$I$ with finite length. The length of the sequence $I$ is denoted
by $|I|$ (for instance in the formula above $|I|=k$). The formal
series (\ref{psi}) defines the functional with a dense domain in
the free Fock space. For instance the domain of each free coherent
state for $\lambda\in (0,\sqrt{p})$ contains the dense space $X$
introduced below.

We define the free coherent state $X_I$ of the form
\begin{equation}\label{indicator}
X_I= \sum_{k=0}^{\infty} \lambda^k \left(\frac{1}{p}
\sum_{i=0}^{p-1}A_i^{\dag}\right)^k \lambda^{|I|} A^{\dag}_I
\Omega+ \sum_{l=1}^{\infty} \lambda^{-l}
\left(\sum_{i=0}^{p-1}A_i\right)^l \lambda^{|I|} A^{\dag}_I \Omega
\end{equation}
The sum on $l$ in fact contains $|I|$ terms. For $\lambda\in
(0,\sqrt{p})$ the coherent state $X_I$ lies in the Hilbert space
(the correspondent functional is bounded).

We denote by $X$ the linear span of free coherent states of the
form (\ref{indicator}) and by $X'$ we denote the space of all the
free coherent states (given by (\ref{psi})).

The following definitions and theorems were proposed in
\cite{coherent1}, \cite{coherent2}, \cite{coherent3}.

\bigskip

\noindent{\bf Definition}\qquad {\sl We define the renormalized
pairing of the spaces $X$ and $X'$ as follows:
\be\label{renormalizedproduct}
(\Psi,\Phi)=\lim_{\lambda\to\sqrt{p}-0}\left(1-\frac{\lambda^2}{p}\right)
\langle \Psi,\Phi\rangle \ee Here $\Psi\in X'$, $\Phi\in X$. }

\bigskip

Note that the coherent states $\Psi$, $\Phi$ defined by
(\ref{psi}), (\ref{indicator}) depend on $\lambda$ and the product
$(\Psi,\Phi)$ does not.

\bigskip

\noindent{\bf Definition}\qquad {\sl We denote $\tilde{\cal F}$
the completion of the space $X$ of the free coherent states with
respect to the norm defined by the renormalized scalar product. }

\bigskip

The space $\tilde{\cal F}$ is a Hilbert space with respect to the
renormalized scalar product.

\bigskip

\noindent {\bf Theorem}\qquad {\sl The space of the free coherent
states \be\label{fcstriple} X \stackrel{i}{\longrightarrow}
\tilde{\cal F} \stackrel{j}{\longrightarrow}  X' \ee is a rigged
Hilbert space. }

\bigskip

Define the characteristic functions of $p$--adic disks
\begin{equation}\label{theta}
\theta_k(x-x_0)=\theta(p^{k}|x-x_0|_p);\quad \theta(t)=0,
t>1;\quad \theta(t)=1, t\le 1.
\end{equation}
Here $x$, $x_0\in Z_p$ lie in the ring of integer $p$--adic
numbers and the function $\theta_k(x-x_0)$ equals to 1 on the disk
$D(x_0,p^{-k})$ of radius $p^{-k}$ with the center in $x_0$ and
equals to 0 outside this disk.

We compare the rigged Hilbert spaces of the free coherent states
(\ref{fcstriple}) and of generalized functions over $p$--adic
disk:
$$
D(Z_p)\stackrel{i'}{\longrightarrow}
L^2(Z_p)\stackrel{j'}{\longrightarrow} D'(Z_p)
$$

\bigskip

\noindent {\bf Theorem}\qquad {\sl The map $\phi$ defined by
$$
\phi:\quad X_I\mapsto p^{|I|}\theta_{|I|}(x-I);
$$
extends to an isomorphism $\phi$ of the rigged Hilbert spaces:

$$
\begin{array}{ccccc}
X & \stackrel{i}{\longrightarrow} & \tilde{\cal F} &
\stackrel{j}{\longrightarrow} & X' \\
\downarrow\lefteqn{\phi}&&\downarrow\lefteqn{\tilde\phi}&&
\downarrow\lefteqn{\phi'}\\
D(Z_p) & \stackrel{i'}{\longrightarrow} & L^2(Z_p) &
\stackrel{j'}{\longrightarrow} & D'(Z_p)
\end{array}
$$
between the rigged Hilbert space of the free coherent states (with
the pairing given by the renormalized scalar product) and the
rigged Hilbert space of generalized functions over $p$--adic disk.

}

\section{The $p$--adic representation of the Cuntz algebra}

In the present section we construct  the representation of the
Cuntz algebra in the space $L^2(Z_p)$ of quadratically integrable
functions on a $p$--adic disk. We will call this representation
the $p$--adic representation. Equivalent representations (without
application of $p$--adic analysis) were considered in \cite{BY}.

Let us define the following operators in $L^2(Z_p)$
\be\label{Adagpadic} A^{\dag}_i \xi(x)=\sqrt{p}\theta_1(x-i)
\xi([\frac{1}{p}x]); \ee \be\label{Apadic} A_i
\xi(x)=\frac{1}{\sqrt{p}}\xi(i+px). \ee Here
$$
[x]=x-x(\hbox{mod }1)
$$
for $x\in Q_p$ is the integer part of $x$. $\theta_1(x-i)$ is an
indicator (or characteristic function) of the $p$-adic disk with
the center in  $i$ and the radius $p^{-1}$.

We have the following

\bigskip

\noindent{\bf Theorem}\qquad{\sl The operators $A_i^{\dag}$ and
$A_i$ defined by (\ref{Adagpadic}) and (\ref{Apadic}) are mutually
adjoint and satisfy the relations of the Cuntz algebra
(\ref{aac}), (\ref{cuntz}):
$$
A_iA_j^{\dag}=\delta_{ij}.
$$
$$
\sum_{i=0}^{p-1}A_i^{\dag} A_i=1.
$$
}

\bigskip

\noindent{\it Proof}\qquad The commutation relations of the
introduced above operators look as follows
$$
A_iA_j^{\dag} \xi(x)=A_i\sqrt{p}\theta_1(x-j) \xi([\frac{1}{p}x])=
\theta_1(i+px-j) \xi([\frac{1}{p}(i+px)])=
$$
$$
=\theta_1(i+px-j)
\xi([\frac{1}{p}i+\frac{1}{p}px])=\delta_{ij}\xi(x);
$$
because $\theta_1(i+px-j)=\delta_{ij}$. Therefore
$$
A_iA_j^{\dag}=\delta_{ij}.
$$
Let us consider
$$
A_i^{\dag} A_i\xi(x)=A_i^{\dag} \frac{1}{\sqrt{p}}\xi(i+px)=
\theta_1(x-i) \xi(i+p[\frac{1}{p}x]).
$$
If $x\ne i (\hbox{mod }  p)$ then  $\theta_1(x-i)=0$.

Therefore the result of application of this operator can be
nonzero only for $x= i (\hbox{mod } p)$. Therefore we can change
the argument of the function $\xi$ by $x$. We get
$$
A_i^{\dag} A_i\xi(x)= \theta_1(x-i) \xi(x);
$$
$$
\sum_{i=0}^{p-1}A_i^{\dag} A_i\xi(x)=  \xi(x);
$$
Therefore
$$
\sum_{i=0}^{p-1}A_i^{\dag} A_i=1.
$$
Let us calculate the adjoint to $A_i^{\dag}$.
$$
\int_{Z_p}A_i^{\dag}\xi(x)\eta(x)d\mu(x)= \int_{Z_p}
\sqrt{p}\theta_1(x-i) \xi([\frac{1}{p}x]) \eta(x)d\mu(x)=
$$
$$
=\sqrt{p}\int_{Z_p} \theta_1(x-i) \xi([\frac{1}{p}(i+x-i)])
\eta(i+x-i)d\mu(x)=
$$
$$
=\sqrt{p}\int_{Z_p} \theta_1(x-i) \xi([\frac{1}{p}(x-i)])
\eta(i+x-i)d\mu(x)=
$$
$$
=\sqrt{p}\int_{||x-i||\le p^{-1}}  \xi(\frac{x-i}{p})
\eta(i+x-i)d\mu(x)=
$$
$$
=\sqrt{p}\int_{||x-i||\le p^{-1}}  \xi(\frac{x-i}{p})
\eta(i+p\frac{x-i}{p} ) p^{-1} d\mu\left(\frac{x-i}{p}\right)=
$$
$$
=\frac{1}{\sqrt{p}}\int_{Z_p}\xi(x)\eta(i+px)d\mu(x)=
\int_{Z_p}\xi(x)A_i\eta(x)d\mu(x)
$$
because for $p$-adic Haar measure $d\mu(px)=p^{-1}d\mu(x)$.

This finishes the proof of the theorem.

\section{The $p$--adic representation as GNS}

Let us define the linear functional $\langle\cdot\rangle$ on the
Boltzmann algebra as follows
\begin{equation}\label{padicstate}
\langle A_{I}^{\dag}A_{J} \rangle =p^{-\frac{1}{2}(|I|+|J|)}
\end{equation}
With $A_{I}^{\dag}$ defined by (\ref{Adag}) and $A_{I}$ adjoint.

In the present section we prove that the considered in the
previous section the $p$--adic representation of the Boltzmann
algebra is unitary equivalent to the GNS representation generated
by the state $\langle\cdot\rangle$.

\bigskip

\noindent {\bf Theorem}\qquad

{\sl

1) The functional $\langle\cdot\rangle$ is a state.

2) In the corresponding GNS representation the condition
(\ref{cuntz}) is satisfied.

3) The corresponding GNS representation is unitarily equivalent to
the representation realized in the space of (quadratically
integrable) functions on $p$-adic disk by the formula
(\ref{Apadic}):
$$
A_i \xi(x)=\frac{1}{\sqrt{p}}\xi(i+px).
$$
}

\bigskip

\noindent {\it Proof}\qquad Let us  prove that for $X$ in the
quantum Boltzmann algebra the functional (\ref{padicstate}) can be
calculated by the integration over $p$-adic variable
\begin{equation}\label{construct}
\langle X\rangle  = (1,X \,1)=\int_{Z_p}(X\, 1) d\mu(x)
\end{equation}
Here $(\cdot,\cdot)$ is the scalar product in the Hilbert space of
square integrable functions on $p$--adic disk $Z_p$ and the action
of $X$ at the RHS is defined by (\ref{Adagpadic}), (\ref{Apadic}).

The formula (\ref{construct}) can be proved by direct calculation.
It is sufficient to prove (\ref{construct})  for monomials
$A^{\dag}_I A_J$. We get
$$
A^{\dag}_I A_J\, 1=A^{\dag}_I p^{-\frac{1}{2}|J|}=
p^{\frac{1}{2}|I|} \theta_{|I|}(x-I) p^{-\frac{1}{2}|J|}
$$
Therefore
$$
\langle A^{\dag}_I A_J \rangle=p^{\frac{1}{2}(|I|-|J|)}
\int_{Z_p}\theta_{|I|}(x-I)  d\mu(x)=p^{\frac{1}{2}(|I|-|J|)}
p^{-|I|}= p^{-\frac{1}{2}(|I|+|J|)}
$$
which gives (\ref{padicstate}).

Since $\langle X\rangle  = (1,X \,1)$ this functional is a state.
The equivalence of GNS and $p$-adic representations  of Boltzmann
algebra follows from  (\ref{construct}) and the fact that $1$ is a
cyclic vector in $p$-adic representations  of Boltzmann algebra.
This finishes the proof of the theorem.

\section{Representation of the Cuntz algebra in the space of FCS}

In the present section we show how to construct non--Fock
representation of the quantum Boltzmann algebra starting from the
Fock representation. We will construct the $p$--adic
representation of the Cuntz algebra by regularization of action of
operators from the quantum Boltzmann algebra on the free coherent
states.

Let us introduce some notations. The free coherent states are
eigenvectors of the annihilation operator $\sum_{i=0}^{p-1}A_i$:
$$
\sum_{i=0}^{p-1}A_i \Phi= \lambda \Phi.
$$
Here $\Phi$ is a function of $\lambda\in (0,\sqrt{p})$. Free
coherent state (FCS) $\Phi$ is given by
$$
\Phi=\sum_I \Phi_I \lambda^{|I|} A^{\dag}_I \Omega.
$$
Here $\Omega$ is the vacuum vector in the free Fock space,
$\Phi_I$ satisfies (\ref{cascade}).

We will also use the notation
$$
\Phi=A^{\dag}_{\Phi} \Omega.
$$

The space of FCS is isomorphic to the space of generalized
functions on a $p$--adic disc. Action of the generalized functions
on test functions is given by the renormalized scalar product on
the space of free coherent states \be\label{renormalized} \left(
\Phi, \Psi \right)=\lim_{\lambda\to\sqrt{p}-0}
\left(1-\frac{\lambda^2}{p}\right)\langle\Phi, \Psi\rangle. \ee

Let us introduce the representation $T$ of the Cuntz algebra in
the space of FCS.

\bigskip

\noindent{\bf Lemma}\qquad  {\sl The regularizations
$T^{\dag}_{i}=T(A^{\dag}_{i})$, $T_{i}=T(A_{i})$ of the right
shift operators on FCS
\begin{equation}\label{cuntzcoh}
T^{\dag}_{i}\Phi =\frac{1}{\sqrt{p}} \left(\lambda A^{\dag}_{\Phi}
A^{\dag}_{i} +\Phi_{\emptyset}\right) \Omega=
\frac{1}{\sqrt{p}}\left(\sum_I  \Phi_I \lambda^{|I|+1}
A^{\dag}_{I} A^{\dag}_{i} +\Phi_{\emptyset}\right) \Omega;
\end{equation}
\be\label{cuntzcoh1} T_{i}\Phi={\sqrt{p}}\sum_I  \Phi_{iI}
\lambda^{|I|}A^{\dag}_{I}\Omega= {\sqrt{p}}
\lambda^{-1}\left(A_{i}A^{\dag op}_{\Phi}\right)^{op}\Omega. \ee
where
$$
\left(A^{\dag}_{i_{k-1}}...A^{\dag}_{i_{0}} \right)^{op}=
A^{\dag}_{i_0}...A^{\dag}_{i_{k-1}}.
$$
map the space of FCS into itself and define the representation of
the Cuntz algebra.}

\bigskip

\noindent{\it Proof}\qquad Check that the introduced operators map
the space of FCS into itself. For the operator $T_{i}$ this
follows from condition (\ref{cascade}) for function $\Phi_{iI}$.
For $T_{i}^{\dag}$ this follows from the formula
\begin{equation}\label{Leibnitz}
\sum_{i=0}^{p-1}A_i  A^{\dag}_{\Phi}= \lambda A^{\dag}_{\Phi} +
\Phi_{\emptyset} \sum_{i=0}^{p-1}A_i.
\end{equation}

Prove that $T_{i}T_{j}^{\dag}=\delta_{ij}$.
$$
T_{i}T_{j}^{\dag} \Phi=T_{i}\frac{1}{\sqrt{p}} \left(\lambda
A^{\dag}_{\Phi} A^{\dag}_{j} +\Phi_{\emptyset}\right) \Omega=
\lambda^{-1}\left(A_{i} \left(\lambda A^{\dag}_{\Phi} A^{\dag}_{j}
+\Phi_{\emptyset}\right)^{op} \right)^{op}\Omega=\delta_{ij}\Phi.
$$
Let us check that $\sum_{i=0}^{p-1}T_{i}^{\dag}T_{i}=1$.
$$
\sum_{i=0}^{p-1}T_{i}^{\dag}T_{i}\Phi=\sum_{i=0}^{p-1}T_{i}^{\dag}
{\sqrt{p}} \lambda^{-1}\left(A_{i}A^{\dag
op}_{\Phi}\right)^{op}\Omega=
$$
$$
=\sum_{i=0}^{p-1} \left( \left(A_{i}A^{\dag op}_{\Phi}\right)^{op}
A^{\dag}_{i} +\Phi_{i}\right) \Omega= \Phi.
$$
Check that in the scalar product $\langle \cdot, \cdot \rangle$
the operators $T_{i}$, $T_{i}^{\dag}$ are adjoint:
$$
\langle T_{i}\Phi,\Psi\rangle = {\sqrt{p}}\sum_I
\Phi_{iI}^{*}\Psi_{I} \lambda^{2|I|};
$$
$$
\langle \Phi,T_{i}^{\dag}\Psi\rangle =\frac{1}{\sqrt{p}}
\left(\Phi^{*}_{\emptyset}\Psi_{\emptyset}+ \lambda^2 \sum_I
\Phi_{iI}^{*}\Psi_{I} \lambda^{2|I|} \right).
$$
This implies that for $\lambda\to\sqrt{p}-0$ one gets $(T_{i}\Phi,
\Psi) =( \Phi, T_{i}^{\dag}\Psi )$.

This finishes the proof of the Lemma.

\bigskip

\noindent{\bf Lemma}\qquad  {\sl The representation of the Cuntz
algebra defined by the operators $T_{i}$, $T_{i}^{\dag}$ is
unitarily equivalent to the $p$--adic representation.}

\bigskip

\noindent{\it Proof}\qquad To prove this it is enough to calculate
the action of the operators $T_{i}$ and $T_{i}^{\dag}$ on the FCS
$X_I$, which correspond to the normed indicators of $p$--adic
disks. We get
$$
T^{\dag}_{i}X_I =\frac{1}{\sqrt{p}} \left(\lambda A^{\dag}_{X_I}
A^{\dag}_{i} +1 \right) \Omega= \frac{1}{\sqrt{p}} X_{iI}.
$$
$$
T_{i}X_I= {\sqrt{p}} \lambda^{-1}\left(A_{i}A^{\dag
op}_{X_I}\right)^{op}\Omega =\sqrt{p} \delta_{ii_0} X_{i_1 \dots
i_{k-1}};
$$
$$
T_{i}X_{\emptyset}=\sqrt{p}  X_{\emptyset};
$$
where $I=i_0 \dots i_{k-1}$.

This finishes the proof of the Lemma.

\bigskip

Let us prove that the $p$--adic representation of the Cuntz
algebra coincides with the restriction of action of antifock
representation $AF$ of the quantum Boltzmann algebra on the space
of FCS.

\bigskip

\noindent{\bf Theorem}\qquad{\sl $p$--Adic representation is the
GNS representation generated by the state on the space of FCS
\begin{equation}\label{afeqtcoh}
\langle X \rangle=(1,X\,1)_{L^2(Z_p)}=\left( \hat 1, AF(X)\, \hat
1 \right)
\end{equation}
Here
$$
\hat 1=\sum_{k=0}^{\infty} \lambda^k \left(\frac{1}{p}
\sum_{i=0}^{p-1}A_i^{\dag}\right)^k \Omega
$$
is the coherent state corresponding to the indicator of $Z_p$ and
$AF$ is the antifock representation. }

\bigskip

\noindent{\it Proof}\qquad From (\ref{cuntzcoh}),
(\ref{cuntzcoh1}) one gets for the action of $T_i$, $T^{\dag}_i$
on free coherent state
$$
T^{\dag}_{i}\Phi=\frac{\lambda}{\sqrt{p}} AF(A^{\dag}_{i})\Phi+
\Phi_{\emptyset} \Omega;
$$
$$
T_{i}\Phi=\frac{\sqrt{p}}{\lambda} AF(A_{i})\Phi.
$$
Then (\ref{renormalized}) implies the Theorem.

\bigskip

\noindent{\bf Remark}\qquad Note that the operators
$AF(A^{\dag}_{i})$, $AF(A_{i})$ do not map the space of FCS into
itself.

\bigskip

\centerline{\bf Acknowledgements.}

The author would like to thank I.V.Volovich, V.A.Avetisov,
A.H.Bikulov and A.Yu.Khrennikov for discussions and valuable
comments. This work has been partly supported by INTAS (grant No.
9900545), by CRDF grant UM1--2421--KV--02 and The Russian
Foundation for Basic Research (project 02-01-01084).

\end{document}